\documentclass[twocolumn]{aastex62}

\newcommand{\be}{\begin{equation}}
\newcommand{\ee}{\end{equation}}
\usepackage{amsmath}
\usepackage{graphicx,bm}
\usepackage[colorinlistoftodos]{todonotes}
\usepackage{float}
\hypersetup{breaklinks,colorlinks,urlcolor=blue,citecolor=blue,linkcolor=blue}

\renewcommand{\arraystretch}{0.5}
\begin{document}

\title{Finding the remnants of the Milky Way's last neutron star mergers}

\author[0000-0003-4960-8706]{Meng-Ru Wu}
\affil{Institute of Physics, Academia Sinica, 
  Taipei, 11529, Taiwan;~\href{mailto:mwu@gate.sinica.edu.tw}{\rm{mwu@gate.sinica.edu.tw}}}
\affil{Institute of Astronomy and Astrophysics, Academia Sinica, 
  Taipei, 10617, Taiwan} 
  
\author[0000-0002-6389-2697]{Projjwal Banerjee}
\affil{Department of Astronomy, School of Physics and Astronomy, Shanghai Jiao Tong University, Shanghai 200240, China;~\href{mailto:projjwal.banerjee@gmail.com}{\rm{projjwal.banerjee@gmail.com}}}
\affil{Department of Physics, Indian Institute of Technology Palakkad, Palakkad, Kerala 678557, India}

\author{Brian~D.~Metzger}
\affil{Department of Physics and Columbia Astrophysics
  Laboratory, Columbia University, Pupin Hall, New York, NY 10027,  USA;~\href{mailto:bdm2129@columbia.edu}{\rm{bdm2129@columbia.edu}}} 

\author[0000-0002-3825-0131]{Gabriel Mart\'inez-Pinedo}
\affil{GSI Helmholtzzentrum f\"ur Schwerionenforschung,
64291 Darmstadt, Germany}
\affil{Institut f{\"u}r Kernphysik
  (Theoriezentrum), Technische Universit{\"a}t Darmstadt,
  64289 Darmstadt, Germany}
  
\author{Tsuguo Aramaki}
\affil{SLAC National Accelerator Laboratory/Kavli Institute for Particle Astrophysics and Cosmology,
Menlo Park, CA 94025, USA}

\author{Eric Burns}
\affil{NASA Postdoctoral Program Fellow, Goddard Space Flight Center, Greenbelt, MD 20771, USA}

\author{Charles J. Hailey}
\affil{Department of Physics and Columbia Astrophysics
  Laboratory, Columbia University, Pupin Hall, New York, NY 10027,  USA} 

\author{Jennifer Barnes}
\altaffiliation{NASA Einstein Fellow}
\affil{Department of Physics and Columbia Astrophysics
  Laboratory, Columbia University, Pupin Hall, New York, NY 10027,  USA} 

\author{Georgia Karagiorgi}
\affil{Department of Physics, Columbia University, Pupin Hall, New York, NY 10027, USA}


\date{\today}

\begin{abstract}
  The discovery of a binary neutron star merger (NSM) through both its
  gravitational wave and electromagnetic emission has revealed these
  events to be key sites of $r$-process nucleosynthesis.  Here, we evaluate the
  prospects of finding the remnants of Galactic NSMs by detecting the
  gamma-ray decay lines from their radioactive $r$-process ejecta.
  We find that $^{126}$Sn, which has several
  lines in the energy range 415--695~keV and resides close to the second
  $r$-process peak, is the most promising isotope, because of its 
half-life $t_{1/2}=2.30(14)\times 10^{5}$~yr being comparable to the ages of recent NSMs.
  Using a Monte Carlo procedure, we predict that multiple
  remnants are detectable as individual sources by next-generation
  $\gamma$-ray telescopes which achieve \mbox{sub-MeV} line sensitivities of
  $\sim 10^{-8}$--$10^{-6}$~$\gamma$~cm$^{-2}$ s$^{-1}$.  However,
  given the unknown locations of the remnants, the most promising
  search strategy is a systematic survey of the Galactic plane and
  bulge extending to high Galactic latitudes.  Individual known
  supernova remnants which may be mis-classified NSM remnants could also be targeted, 
  especially those located outside the Galactic plane.
  Detection of a moderate sample of Galactic NSM remnants would provide important clues to
  unresolved issues such as the production of actinides in NSMs, properties of merging NS binaries, and even help distinguish them from rare supernovae as current Galactic $r$-process sources.
  We also investigate the diffuse flux from longer-lived nuclei (e.g.~$^{182}$Hf)
  that could in principle trace the
  Galactic spatial distribution of NSMs over longer timescales, but find that the
  detection of the diffuse flux appears challenging even with
  next-generation telescopes.
\end{abstract}

\keywords{gamma-ray astronomy, r process}

\section{Introduction}

Roughly half of the naturally occurring isotopes heavier than the iron group are
created through the process of rapid neutron capture 
($r$-process;~\citealt{Burbidge+57,Cameron57}; see~\citealt{Cowan+19} for 
a recent review).  
Although the basic physical conditions needed for the $r$-process 
are well understood (e.g.~\citealt{Hoffman+97}), the astrophysical 
site or sites giving rise to the requisite high neutron flux remains debated.  
Among the primary candidates are core collapse supernovae 
(SNe; e.g.~\citealt{Meyer+92,Takahashi+94,Woosley+94}) and the 
coalescence of compact neutron star binaries \citep{Lattimer&Schramm74,Symbalisty&Schramm82,Eichler+89,Korobkin+12}.  
For SNe, one can further distinguish the neutrino-driven winds 
from proto-neutron stars (common to most SNe; e.g.~\citealt{Qian&Woosley96,Thompson+01}), 
from the rarer subset of collapse events which give birth to 
rapidly-spinning highly-magnetized neutron stars \citep{Thompson+04,Metzger+07,Winteler+12,Mosta+18} 
or hyper-accreting black holes \citep{Fryer+06,Siegel+18}.

Our understanding of the $r$-process advanced dramatically following 
the discovery of a binary neutron star merger (NSM) through both its 
gravitational waves \citep{LIGO+17DISCOVERY} and electromagnetic 
light \citep{LIGO+17CAPSTONE}.  
This event, dubbed GW170817, was accompanied by fading visual and 
infrared emission (e.g.~\citealt{Coulter+17,Soares-Santos+17}), 
which was widely interpreted as being powered by the radioactive 
decay of freshly synthesized $r$-process nuclei \citep{Li&Paczynski98,Metzger+10,Barnes&Kasen13,Tanaka&Hotokezaka13}.  
Modeling of the light curve indicates a total $r$-process ejecta mass 
of $\approx 0.03$--0.06~M$_{\odot}$ 
(e.g.~\citealt{Drout:2017ijr,Cowperthwaite+17,Kasen+17,Villar+17,Kawaguchi+18,Wanajo18,Wu+19}).  
The large quantity of ejecta from GW170817 shows that NSMs are major, 
if not dominant, sources of the Galactic $r$-process (e.g.~\citealt{Kasen+17}).  
The mergers of compact binaries comprised of a neutron star (NS) and stellar-mass 
black hole (BH) can also eject large quantities of neutron-rich 
$r$-process material, provided that the BH is rapidly spinning and 
of sufficiently low mass to tidally disrupt the NS before the latter 
plunges inside the BH event horizon (e.g.~\citealt{Foucart+18}).

Despite this progress, a few key questions remain open.  
For instance, it is unclear whether NSMs occur sufficiently promptly 
following the first generations of star formation in the universe to 
explain the high $r$-process abundance in metal-poor halo 
stars \citep{Sneden+08} and dwarf galaxies \citep{Ji+16} (see e.g.,~\citealt{vandeVoort2015, Hirai2015,Shen+15,Wehmeyer2015,Cote+18, Safarzadeh:2018fdy}).  
Studies of Galactic chemical evolution also indicate that the growth in 
the abundances of Europium relative to $\alpha$-process elements 
(which originate mainly from SNe) points to an $r$-process source   
which tracks ongoing star formation \citep{Cote+18,Hotokezaka:2018aui} 
instead of the delayed population generally predicted for NSMs 
(however, see \citealt{Beniamini&Piran19}).  
Separately, it is not clear whether the observed diversity 
in the abundance patterns of individual Galactic $r$-process 
events \citep{Honda+06,Holmbeck+18,Ji:2018_2} is an indication 
of separate production sites (e.g. NS-NS versus NS-BH mergers 
versus rare SNe), or diversity within an underlying similar event.  
While the late-time infrared emission from GW170817 provides evidence 
for the production of lanthanide elements (atomic number $A \gtrsim 140$; e.g.~\citealt{Chornock+17,Tanvir+17,Pian+17}), no strong evidence 
exists for the production of heavier nuclei near the third 
$r$-process peak ($A \gtrsim 195$).

Freshly synthesized $r$-process elements are radioactive.  Gamma-rays
\citep{Qian+98,Qian+99} or X-rays \citep{Ripley+14} from decaying
$r$-process nuclei in young supernova (SN) remnants can therefore
provide direct evidence of their production.  Decay lines from the
(non $r$-process) isotope $^{44}$Ti were detected from the SN remnants
Cas A \citep{Iyudin1994,Vink+01,Renaud+06,Boggs:2015xx,Siegert:2015xx,Grefenstette+17} and SN 1987A \citep{Grebenev+12}.  
The biggest challenge to detect $r$-process elements in SNe remnants is
that their expected abundances are many orders of magnitude smaller
than lighter nuclei, assuming that the $r$-process isotopes are
produced in equal quantity in all SNe and that the latter are major
contributors to the total Galactic $r$-process abundances.  If instead
only a small subset of SNe produce the $r$-process and the per-event
yields are higher, then this subset of SNe would have higher line
fluxes (but a large number of remnants must then be searched to
discover even one $r$-process source).

Because of their guaranteed larger $r$-process yields, the remnants of
past NSMs in our Galaxy may be more promising $\gamma$-ray line sources
\citep{Ripley+14}.  NSMs occur in the present-day Milky Way (MW) at an
estimated rate of $f_{\text{NSM}} \sim 10$--100~Myr$^{-1}$, a range which
is consistent with both studies of the Galactic double neutron star
population (e.g.~\citealt{Kim+10}), limits from the LIGO O1/O2 observing
runs \citep{LIGO+18CATALOG}, and constraints based on the $r$-process
yield from GW170817 for total Galactic abundances.  The youngest NSM
remnant in our Galaxy is therefore of age $\sim 10^{4}$--$10^{5}$~yr, a
range fortuitously comparable to the half-lives of several promising
$r$-process isotopes, particularly $^{126}$Sn
($t_{1/2} = 2.30(14)\times 10^{5}$ yr; see Table~\ref{tab:nuc}).

The kinetic energy of the kilonova ejecta from GW170817 was inferred
to be $\sim 10^{51}$~erg, similar to that of SNe \citep[e.g.][]{Villar+17}.  If representative,
then the physical size of NSM remnants, following their shock
interaction with the interstellar medium (ISM), is similar to those of
SN remnants of the same age~\citep{Montes:2016xkx}.  However, since the spatial locations of
the NSM remnants (which are far-outnumbered by SN remnants) are not
known, detecting their $\gamma$-ray line signal may require a creative
search strategy, such as a systematic search of known remnants which
in rare cases might be 
mis-classified as SNe, or a wide-field survey of
the Galactic plane/bulge.  Unlike core collapse SNe, which largely
take place in the high-density Galactic plane, NSMs can take place
with large physical offsets from their birth locations due to NS natal
kicks (e.g.~\citealt{Bloom:1998zi}), in which case their spatial
distribution may extend to higher Galactic latitudes.

As with SNe, old NSM remnants of age $\gg 10^{6}$~yr will eventually
have their material mixed into the ISM of the Galactic halo or disk.
Decay lines from long-lived nuclei (e.g.~$^{182}$Hf, with
$t_{1/2} =  8.90(9)\times 10^{6}$~yr) may thus present as a diffuse
$\gamma$-ray line flux from the Galactic plane, much in the way that
$^{26}$Al ($t_{1/2}=7.17(24)\times 10^{5}$~yr) and $^{60}$Fe
($t_{1/2} = 2.62(4)\times 10^{6}$~yr) are measured in the inner
portions of the MW \citep{Pluschke+01,Smith04,Diehl+06,Wang+07}.

In this paper, we predict the properties of $r$-process $\gamma$-ray line
sources from Galactic NSMs, lay out strategies to detect them, and
highlight the scientific returns of such discoveries.  We provide the
expected number, distances, sky-positions, and angular sizes of
sources, as well as their $\gamma$-ray line fluxes.  
For individual remnants, we focus on the lines from $^{126}$Sn, 
given its optimal half-life and large fluxes.  
Being a nucleus just below the second $r$-process peak,
which does not require particularly neutron-rich ejecta for its
creation, $^{126}$Sn production in NSMs has the benefit of likely
being guaranteed and more robust.  
While less robustly produced, we also consider decay
lines from $^{230}$Th ($t_{1/2} = 7.54(3)\times 10^{4}$~yr), which
would probe the presently-unconstrained production of actinides in
NSMs.  Finally, we estimate the diffuse background from $^{182}$Hf,
under the assumption that, while individual NSM remnants may have
dissolved into the ISM prior to its decay, the angular 
distribution of $^{182}$Hf off the Galactic
plane could distinguish different $r$-process sources.

The line sensitivities of past or existing MeV $\gamma$-ray satellites, 
such as COMPTEL \citep{Schoenfelder+93} or INTEGRAL \citep{Diehl13} 
of $\sim 10^{-5}$ $\gamma$ cm$^{-2}$ s$^{-1}$, are probably not 
sufficient to detect $r$-process lines from NSM remnants. 
However, proposed next generation balloon or satellite 
missions could achieve line sensitivities of 
$\sim 10^{-8}$--$10^{-6}$~$\gamma$~cm$^{-2}$ s$^{-1}$ \citep[see a summary in e.g.,][]{Fryer+19}.
These include balloon missions such as
COSI~\citep{Kierans:2017bmv} and
GRAMS \citep{Aramaki+19}, as well as 
several satellite missions, e.g., AMEGO~\citep{Moiseev:2017mxg},
e-ASTROGAM~\citep{DeAngelis:2017gra},
ETCC~\citep{Tanimori:2017ihu}, HEX-P~\citep{HEXP_Madsen18}, and
LOX~\citep{LOX_Miller2018}.

Rather than addressing the individual prospects of these concepts 
for NSM remnant science, which will be heavily dependent on 
particulars such as the instrument and astrophysical backgrounds, 
our chief goal with this work is to highlight the key multi-messenger 
science and to provide concrete predictions to motivate these concept studies.

\section{NSM remnant distribution}\label{sec:remdis}

\subsection{Spatial Position, Distance, and Age}
 
We model the spatial and age distribution of young (age $\lesssim$ 50~Myr) 
NSM remnants in our Galaxy using the following prescription.
\begin{itemize}
\item We assume that the ``birth'' places of the binary NS systems
  trace the stellar mass distribution of the MW bulge and disk.  We
  model the latter following \citet{mcmillan2017}, particularly
  their eqs.~(1)-(3), using the best-fit values from their Table~3 for
  the MW stellar density.  In reality, binary NS systems are formed at
  essentially the same times and places as the stars themselves.
  However, neglecting SN kicks, the substantial delay due to the slow
  process of gravitational wave-inspiral should result in
  currently-merging binaries tracing the locations of older stars more
  faithfully than that of current star-formation.
\item Neutron stars can receive substantial kicks at birth of up to
  several hundred km s$^{-1}$, which affect the locations of NSM
  events (e.g.~\citealt{Bloom:1998zi}).  From ``birth'' to merger, we account
  for this effect by allowing the binary systems to undergo spatial
  drift from the stellar population, using the observed offset
  distribution of short-duration gamma-ray bursts (GRB) from their
  host galaxies from \citet{fong2013}.  We assume the drift
  directions are isotropically distributed from the birth sites.

  Since some short GRBs occur in galaxies with different properties
  than the spiral-type MW (e.g.~elliptical or S0 galaxies), we
  consider two models for the offset distribution in order to test
  the sensitivity of our results to our assumptions.  Model I uses an
  offset distribution from Fig.~6 in \citet{fong2013} which has
  been normalized to the effective radius $r_{\rm e}$ of the host
  galaxy, taking $r_{\rm e}=7.5$~kpc for the MW.  Model II instead
  uses directly the offset distribution in physical units from Fig.~5
  in \citet{fong2013}.
\item We consider two values for the current NSM rate in the MW,
  $f_{\rm NSM}=10$~Myr$^{-1}$ and 100~Myr$^{-1}$.  The higher of these
  rates is near the mean value inferred from the LIGO O1/O2 runs
  \citep{LIGO+18CATALOG} while the lower rate represents a
  ``conservative" scenario on the very low end of the allowed rate.
  We assume that mergers are distributed uniformly in time, which is
  justified for young merger remnants with ages~$\lesssim 100$ Myr
  much less than the timescale over which the MW star formation rate
  is currently evolving.
\item We perform Monte Carlo sampling over the above temporal and
  spatial distributions for time spans of 10 Myr for the individual source
  detection of $\gamma$-lines from shorter-lived nuclei
  ($t_{1/2}\ll 1$~Myr), and 50 Myr for the diffuse lines from
  longer-lived nuclei ($t_{1/2}\sim 10$~Myr).
\end{itemize}

\begin{figure}[htbp!]
  \centering 
  \includegraphics[width=\linewidth]{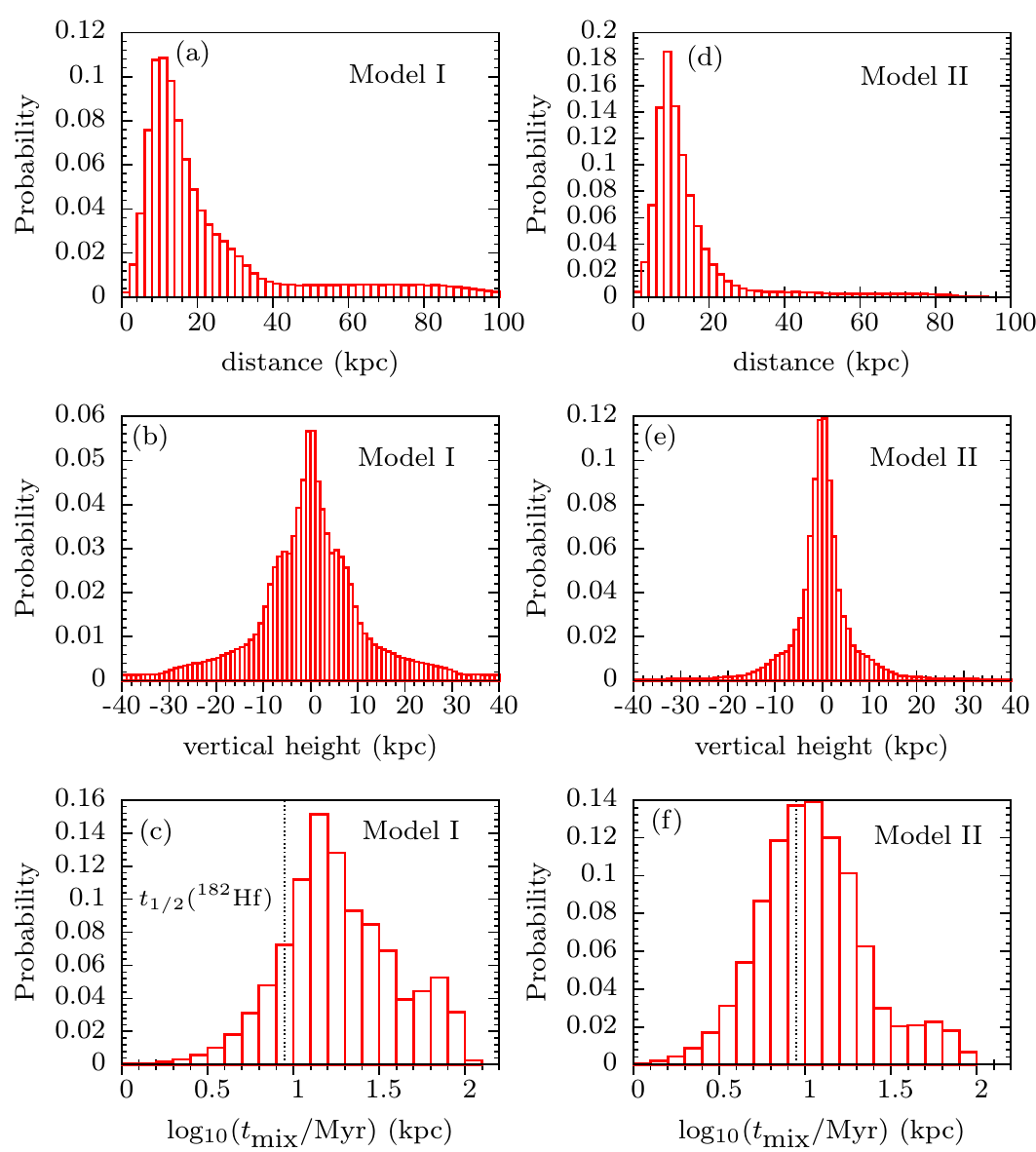}
  \caption{Probability distribution of distances, vertical heights 
  (as measured out of the Galactic plane), and ISM mixing time 
  $t_{\rm mix}$ for a simulated population of Galactic NSM remnants.  
  Results are shown separately for Model I [panels (a)--(c)] and 
  Model II [panels (d)--(f)] for the physical offset of NSMs from the 
  stars (see text).  In the bottom panels (c) and (f), the half-life 
  of $^{182}$Hf is shown for comparison with a vertical dotted line.
  \label{fig:distancePDF_ModelI}}
\end{figure}

The top two rows of Fig.~\ref{fig:distancePDF_ModelI} show the probability
distributions of the distances (from Earth) of the NSM remnants and
their vertical height $z$ off the mid-plane of the MW disk, shown
separately for Model I (left panels) and Model II (right panels).  The
distance distribution in both Models peaks at $\sim 10$~kpc and
smoothly extends to large distances $\sim 100$~kpc.  However, as
expected, the width of the distribution in Model I is larger than that
in Model II due to the assumed offsets, giving rise to a lower
probability of closer remnants in Model I.
The distribution of vertical heights is centered about the midplane ($z = 0$) 
but extending to up to $|z| \sim$ several tens of kpc.  
The height distribution of Model I is again broader than Model II, 
but in both cases a significant fraction $\sim 20$--40\% of the 
remnants are located within $|z| \lesssim 2$~kpc.  

\subsection{Physical and Angular Size}
Following \citet{cioffi1988} for the interaction of SN ejecta with 
the ISM, there are three distinct phases in the evolution of NSM 
remnants before it merges with the ISM.  The initial ``free expansion'' 
phase takes place until the swept-up ISM mass equals that of the 
ejecta, $M_{\rm ej}$. This occurs after a time
\begin{equation}
    t_{\rm sw}= 69.5\left(\frac{0.1 c}{v_{\rm ej}}\right)\left(\frac{M_{\rm ej}}{M_\odot}\right)^{1/3}\left(\frac{n}{{\rm cm^{-3}}}\right)^{-1/3} {\rm \ yr},
\end{equation}
where $v_{\rm ej}$ is the ejecta velocity and $n \approx \rho/m_p$ is 
the ISM particle density (here $\rho$ and $m_p$ are the mass density 
and proton mass, respectively).  After a time $t \gtrsim t_{\rm sw}$, 
the remnant evolves as an energy-conserving Sedov-Taylor (ST) blast 
wave, until radiative cooling becomes important after a time 
\begin{equation}
    t_{\rm PDS}= t_{\rm sw}+1.33\times10^4\left(\frac{E}{10^{51} {\rm \ ergs }}\right)^{3/14} \left(\frac{n}{{\rm cm^{-3}}}\right)^{-4/7} {\rm \ yr,}
\end{equation}
where $E=M_{\rm ej}v_{\rm ej}^2/2$ is the ejecta kinetic energy.  
At times $t \gtrsim t_{\rm PDS},$ the pressure of the hot interior 
of the remnant drives the expansion that is further aided by its 
momentum in a phase known as the ``pressure-driven snowplow (PDS)".  
Finally, the ejecta merges/mixes with the ISM after a 
time\footnote{At the end of the PDS phase, when most of the interior 
thermal energy has been depleted due to radiative cooling, the  
interior pressure becomes negligible. At this point, the remnant 
enters into the so-called ``momentum-conserving snowplow (MCS)" phase, 
where the expansion is solely driven by the momentum of the remnant. 
However, as noted by \citet{cioffi1988}, the remnant usually merges 
with the ISM while still in the PDS or ST phase well before MCS phase 
is reached. Therefore, we neglect the MCS phase in this work.}
\begin{equation}
    t_{\rm mix}=56.84\, t_{\rm PDS}\left(\frac{E}{10^{51} {\rm \ ergs }}\right)^{5/49} \left(\frac{n}{{\rm cm^{-3}}}\right)^{10/49}.
\end{equation}
Combining the above results, the radius of a NSM remnant of age $t$ 
is given by
\begin{eqnarray}
r_{\rm NSM}(t) &=& v_{\rm ej}t   , t\leq t_{\rm sw} \nonumber\\
               &=&R_{\rm sw} +\left(\frac{2.026E(t-t_{\rm sw})^2}{\rho}\right)^{1/5}, t_{\rm sw}<t\leq t_{\rm PDS} \nonumber\\
               &=&R_{\rm PDS}\left(\frac{4}{3}\frac{t}{t_{\rm PDS}}-\frac{1}{3}  \right)^{3/10}  , t_{\rm PDS}< t\leq t_{\rm mix} \nonumber\\
               &=&R_{\rm PDS}\left(\frac{4}{3}\frac{t_{\rm mix}}{t_{\rm PDS}}-\frac{1}{3}  \right)^{3/10}, t>t_{\rm mix}, 
\end{eqnarray}
where $R_{\rm sw}$ and $R_{\rm PDS}$ are the values of $r_{\rm NSM}$ 
at $t_{\rm sw}$ and $t_{\rm PDS}$, respectively.  
The expansion velocity of the remnant can then be estimated 
as $v_{\rm exp}=dr_{\rm NSM}/dt$. 

To estimate the local value of the ISM density at the location 
of each NSM remnant, we assume that gas density drops with 
Galactic radius $r$ and vertical distance $z$ above the MW disk 
according to the following profile from \citet{miller2013}:
\begin{equation}
    n(r)=n_0[1 +(r/R_c)^2 +(z/z_c)^2]^{-3\beta/2}
    \label{eq:miller2013}
\end{equation}
where we take $n_0=0.46$~cm$^{-3}$, 
$R_c=0.42$~kpc, $z_c=0.26$~kpc, and $\beta=0.71$ from the best-fit 
values of \citet{miller2013}.  The angular size $\alpha$ of the 
remnant diameter in radians is then given by
\begin{equation}
    \alpha=\arctan\left(\frac{2r_{\rm NSM}}{d}\right),
\end{equation}
where $d$ is the remnant distance.  

The bottom panels of Fig.~\ref{fig:distancePDF_ModelI} show the
distribution of $t_{\rm mix}$ for all of the NSM remnants in our Monte
Carlo sample, assuming ejecta parameters $E=10^{51}$~erg and
$M_{\rm ej}=0.04$~$M_\odot$, motivated by observations of GW170817.  A
minimum value of $t_{\rm mix} \sim 1$ Myr is reached near the densest
regions in the Galactic Center (GC), but the distribution peaks at
$\sim 10-15$ Myr and extends to larger values $\sim 100$ Myr.  As the
mixing times are much longer than the half-lives of several of the
isotopes of greatest interest (e.g.~$^{126}$Sn, $^{230}$Th), these
$\gamma$-ray lines should still be found as singularly associated with
individual NSM remnant (Sec.~\ref{sec:pointsrc}).  By contrast, the
mixing time can be comparable to the half-life of $^{182}$Hf (shown as
a vertical dashed line in Fig.~\ref{fig:distancePDF_ModelI}),
indicating that this isotope might be substantially mixed with the
ISM, in which case it would instead form a more diffuse $\gamma$-ray background
(Sec.~\ref{sec:diffuse}).

\section{Individual Remnants}\label{sec:pointsrc}

There are only 22 known X- and $\gamma$-ray emitting radioactive
$r$-process nuclei with half-lives ($t_{1/2}$) in the range 
$10^2-10^8$ years. Table~\ref{tab:nuc} in the Appendix lists 
their decay sequences, half-lives, and the major X- and $\gamma$-ray 
line energies and intensities (probability of emitting
a $\gamma$ per decay). 
Figure~\ref{fig:lines} in the Appendix shows an example of 
the X-ray/$\gamma$-ray line spectrum of a remnant of age 
$\approx 5\times 10^{4}$ yr and distance of $9$ kpc, similar 
to those of the youngest Galactic NSM remnants. 

Among these isotopes, $^{126}$Sn, which resides close to the second
$r$-process peak, with $t_{1/2}=2.3\times 10^5$~yr, is the most
promising candidate for NSM remnant $\gamma$-ray searches. First, the
decay sequence of $^{126}\text{Sn} \rightarrow {}
^{126}\text{Sb} \rightarrow{} ^{126}$Te produces a few strong lines with
energy (intensity) of 414.7~keV (98\%), 666.3~keV (100\%), 695.0~keV
(97\%)~\citep{Orth.Dropesky.Freeman:1971,Bargholtz.Becker.ea:1975,Smith.Bunker.ea:1976}\footnote{Note
    that the decay of $^{126}$Sn populates the first excited isomeric
    state of $^{126}$Sb ($J^\pi=5^+$) before reaching its ground state
    ($J^\pi=8^-$).  This isomeric state has a branching ratio of
    86(4)\% by $\beta$-decay to $^{126}$Te, and 14(4)\% by
    isomeric-transition to the ground state of
    $^{126}$Sb~\citep{Orth.Dropesky.Freeman:1971}, which further
    $\beta$-decays to $^{126}$Te.  Therefore, the decay line
    intensities from $^{126}$Sb are a linear-superposition of these two
    sub-channels.  For example, a 720.7~keV line has a $53.80(24)\%$
    intensity~\citep{Bargholtz.Becker.ea:1975} via the latter channel
    but will not be produced via the former, which results in a total
    intensity of only $\approx 7\%$.}  Second, the production of
nuclei near the second peak, like $^{126}$Sn, is almost guaranteed in
every NSM.  Motivated thus, we evaluate the detection prospect of
$^{126}$Sn decay $\gamma$-rays in Sec.~\ref{sec:sn126} and then
discuss the possibility of co-detecting other lines from the actinide
decay in Sec.~\ref{sec:actinides}.

\subsection{$^{126}$\textup{Sn} Individual Sources}\label{sec:sn126}
We assume that each NSM produces an ejecta mass $M_{\rm ej}=0.04~M_\odot$,
consistent with the inferred production yield of GW170817, and contains a
distribution of $r$-process nuclei following the Solar $r$ abundances in 
the mass range $A=90-205$ taken 
from~\citet{Sneden+08}\footnote{The Solar $r$ abundances 
in~\citet{Sneden+08} are given for elements with charge 
number $Z\geq 69$. However, we do not include those between 
$69\leq A\leq 89$ because their solar $r$ abundances may have received
large contribution from other sites like core collapse SNe. 
Moreover, the observed metal-poor star abundances suggests that nuclei
around $A\sim 70$ are likely not co-produced by NSMs~\citep{Cowan:2005eh}.
}. 
This gives a corresponding number fraction per nucleon, or abundance, of
$^{126}$Sn at production, $Y_0=1.7\times 10^{-4}$.

\begin{figure}[htbp!]
  \centering \includegraphics[width=0.9\linewidth]{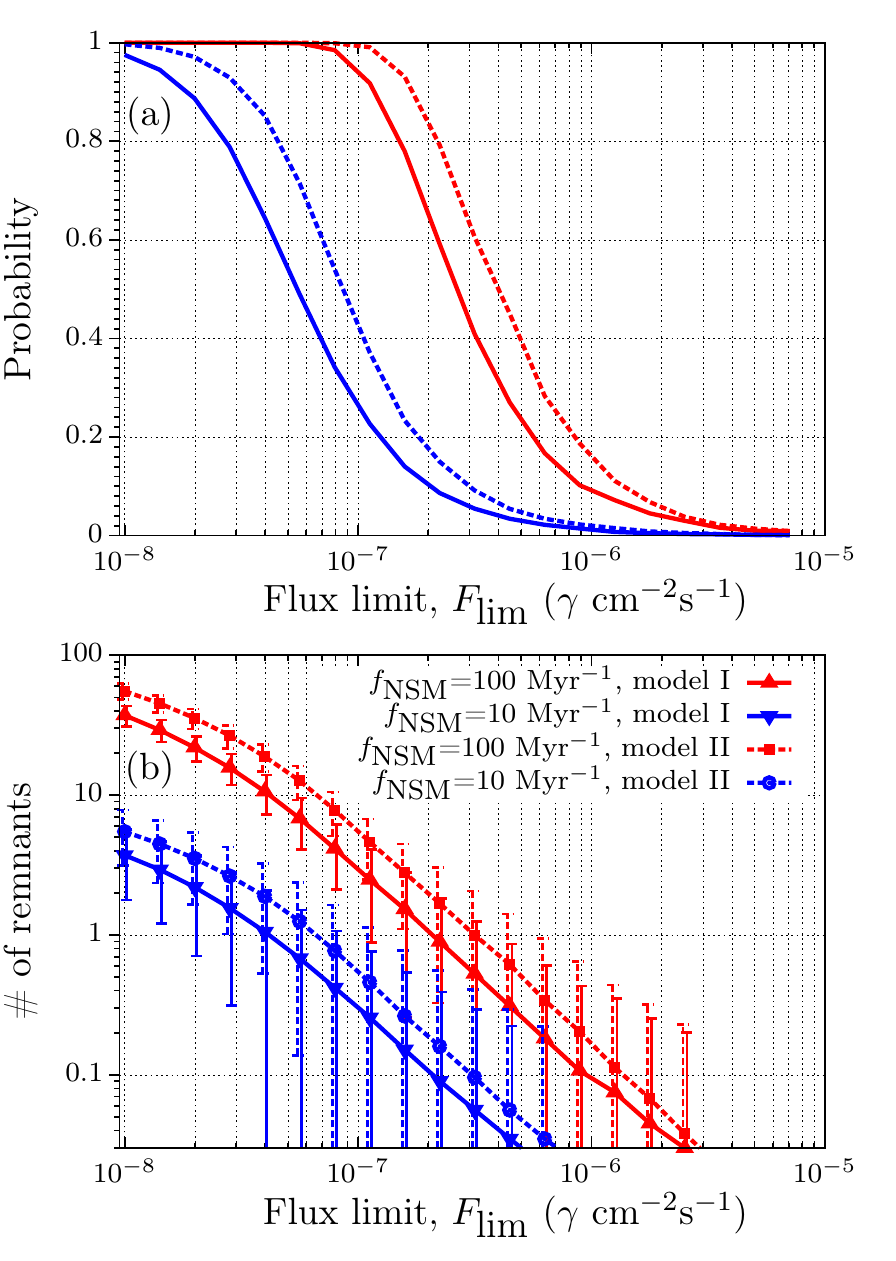}
  \caption{Panel (a): Probability of the existence of at least one NSM 
  remnant with a $^{126}$Sn $\gamma$-ray line flux exceeding a given flux 
  limit, $F_{\rm lim}$.  We show separately the results for two assumptions 
  for the spatial offset of NSMs (Model I and Model II), and for two merger 
  rates, $f_{\rm NSM}=$10~Myr$^{-1}$ and 100~Myr$^{-1}$.  
  Panel (b): Number of remnants with $\gamma$-ray line flux $\ge F_{\rm lim}$ 
  as a function of $F_{\rm lim}$ for the same scenarios as in the top panel.
  Thick curves show the detected number averaged over all realizations, 
  while vertical error bars show the $\pm 1\sigma$ sample variance.
  \label{fig:prob}}
\end{figure}

We then generate $10^{3}$ and $10^{4}$ realizations for the assumed NSM
frequency $f_{\rm NSM}=100$~Myr$^{-1}$ and $10$~Myr$^{-1}$, respectively,
following the method described in Sec.~\ref{sec:remdis} for both
Model I and II.
For each remnant, we calculate the photon number flux $F(d,t)$ with 
an intensity $I_g=100\%$, which corresponds to the intensity of 
the strongest lines at $666.3$~keV from the decay of $^{126}$Sn:
\begin{equation}
F(d,t)=\frac{M_{\rm ej} Y_0 I_g}{4\pi d^2 m_u \tau_0}e^{-t/\tau_0},
\end{equation}
where $m_u$ is the atomic mass unit and 
$\tau_0=2.3\times 10^5\textrm{~yr}/\ln(2)$ is the lifetime of $^{126}$Sn.  
For a given threshold flux limit $F_{\rm lim}$, 
we then calculate the number of NSM remnants having
$F(d,t)> F_{\rm lim}$ for all realizations, separately in each 
model and for both values of $f_{\rm NSM}$.

Figure~\ref{fig:prob} shows the probability of having \emph{at least one}
remnant (top panel), as well as the expected number of remnants (bottom panel),
for which the $^{126}$Sn line flux exceeds a given limiting value 
$F_{\rm lim}$.  
Our results do not depend sensitively on the chosen model of the remnant 
offset distribution.  Compared to Model I, Model II yields a slightly 
higher probability because its predicted smaller offset distribution 
leads to mergers being on average closer (Fig.~\ref{fig:distancePDF_ModelI}).

The detection probability does, however, depend sensitively on the 
assumed NSM rate, $f_{\rm NSM}$. 
For the value $f_{\rm NSM}=100$~Myr$^{-1}$ in the middle of the range 
allowed by the LIGO discovery of GW170817, we find that $\sim 1$--$5$ 
remnants have line fluxes above 
$F_{\rm lim}\sim 10^{-7}$~$\gamma$~cm$^{-2}$~s$^{-1}$, with a 
$\gtrsim 90\%$ probability of at least one remnant above this threshold.  
For a lower detection threshold 
$F_{\rm lim}\sim 10^{-8}$~$\gamma$~cm$^{-2}$~s$^{-1}$, 
the prospects are obviously better.  
Even for the most conservative merger rate $f_{\rm NSM}=10$~Myr$^{-1}$,  
the probability of having more than one merger remnant with line flux 
above $F_{\rm lim}$ is as large as $96\%$.
For the high merger rate $f_{\rm NSM}=100$~Myr$^{-1}$, 
there should exist $\gtrsim 30$ NSM remnants above
$10^{-8}$~$\gamma$~cm$^{-2}$~s$^{-1}$.

\begin{figure*}[htbp!]
  \centering 
  \includegraphics[width=0.6\textwidth]{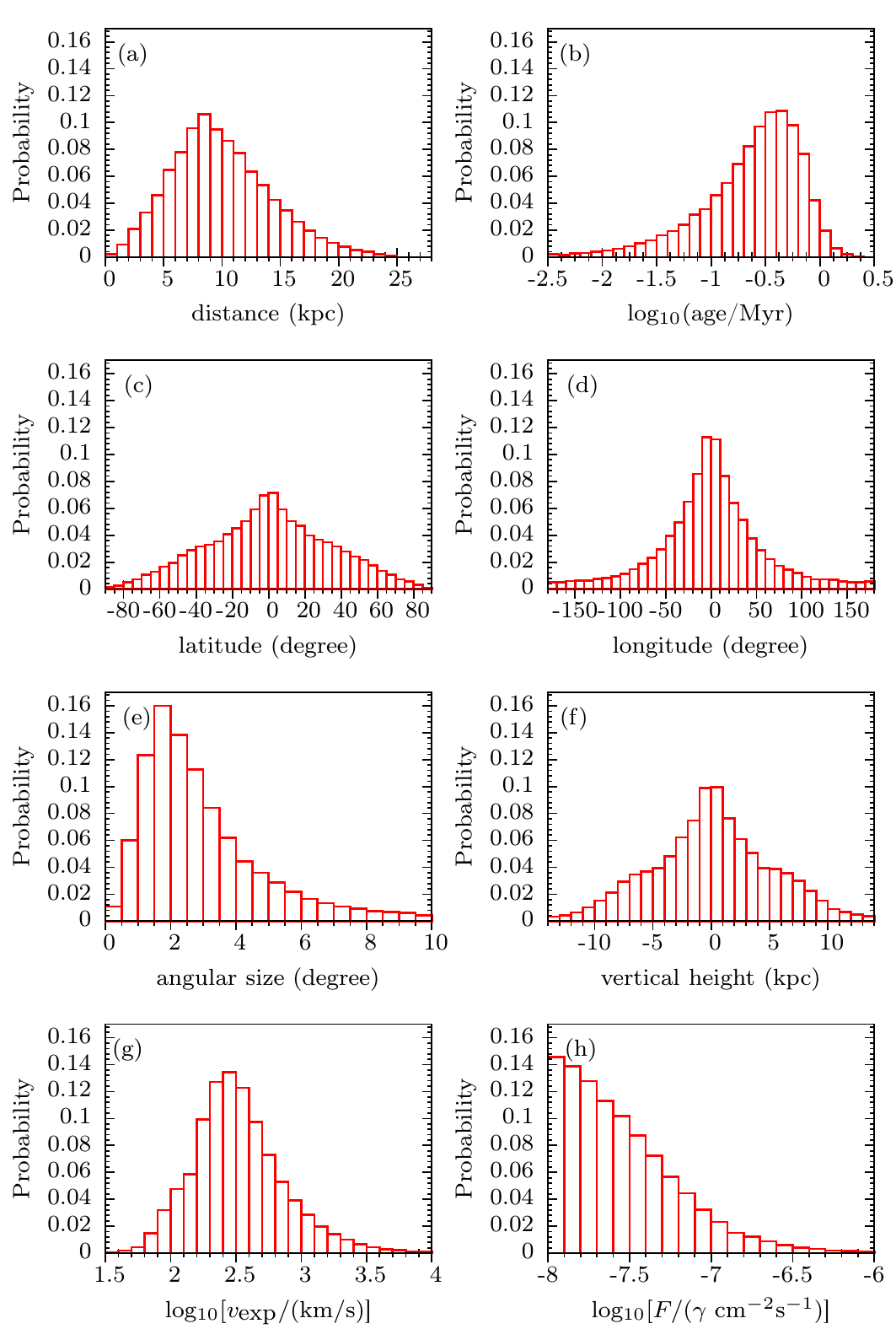}
  \caption{Probability distributions of several key properties of NSM 
  remnants with $^{126}$Sn $\gamma$-ray line fluxes 
  $F\geq 10^{-8}$~$\gamma~$cm$^{-2}$ s$^{-1}$.  
  Properties shown include distance, age, latitude, longitude 
  (in the Galactic coordinate), angular size, vertical height above 
  the MW disk, remnant expansion velocity $v_{\rm exp}$, and 
  line flux $F(d,t)$ at 666.3~keV.  
  We use Model I for the NSM spatial offsets and assume a Galactic NSM 
  rate of $f_{\rm NSM}=100$ Myr$^{-1}$.  For all remnants, we have assumed 
  an $r$-process ejecta mass $M_{\rm ej}=0.04 M_\odot$ with 
  $Y(^{126}{\rm Sn})=1.7\times 10^{-4}$.  
  }\label{fig:dist_detrem}
\end{figure*}

Figure~\ref{fig:dist_detrem} further shows distribution of distances, ages,
Galactic latitudes, longitudes, angular sizes, vertical heights, expansion
velocities $v_{\rm exp}$, and $\gamma$-ray line flux $F(d,t)$, for 
all remnants satisfying $F(d,t)>F_{\rm lim}=10^{-8}$~$\gamma$~cm$^{-2}$~s$^{-1}$ (for Model I with $f_{\rm NSM}=100$~Myr$^{-1}$).  
The detectable remnant population are mostly younger than $1$~Myr and 
located at characteristic distances of $\sim 9$~kpc.  Most have angular 
sizes of $\sim 2^\circ$ and are within $\sim\pm 5$~kpc of Galactic plane.  
Most of the remnants have low expansion velocities 
$v_{\rm exp}<3\times 10^{3}$~km~s$^{-1}$, indicating narrow predicted 
$\gamma$-ray line widths (Sec.~\ref{sec:survey}).

\begin{figure*}[htbp!]
  \centering 
  \includegraphics[width=0.9\textwidth]{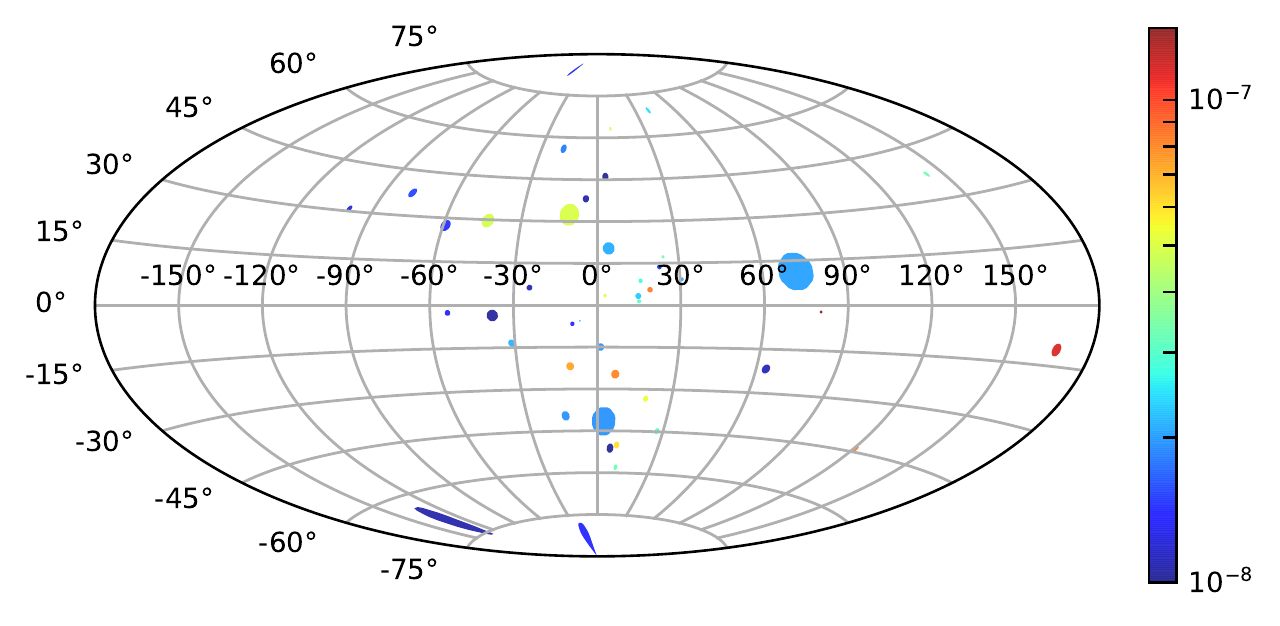}
  \caption{Example realization of the sky map of individual NSM remnant 
  $\gamma$-ray sources with line fluxes 
  $F>F_{\rm lim}=10^{-8}~\gamma$~cm$^{-2}$~s$^{-1}$ for the same 
  scenario in Fig.~\ref{fig:dist_detrem}. 
  The cicular/elliptical shaded areas represent the
  angular sizes of the remnants and the colors represent the emitted 666.3~keV line fluxes in units of
  $\gamma$~${\rm cm}^{-2}{\rm s}^{-1}$, indicated on the legend.  
  The center of the plot marks the direction of the GC.\label{fig:map_ps}}
\end{figure*}

Interestingly, roughly $40\%$ of the remnants reside within $\pm 20^\circ$ 
from the GC.  Furthermore, $\sim 4\%$ and $17\%$ of the NSM remnants are 
located inside $\sim$ the Galactic bulge defined by a sphere with a radius 
of $2$~kpc and the galactic disk defined by a typical scale height of 
$1$~kpc, respectively.  As we discuss in Sect.~\ref{sec:search}, this 
motivates a survey strategy focused on the Galactic plane and bulge. 
Figure~\ref{fig:map_ps} shows one realization (from Model I, with 
$f_{\rm NSM}=100$~Myr$^{-1}$) of the sky map of spatial positions and 
$\gamma$-ray fluxes of remnants obeying 
$F(d,t)>F_{\rm lim}=10^{-8}$~$\gamma$~cm$^{-2}$~s$^{-1}$.  
The distances and vertical heights of the detectable population of NSM 
remnants are usefully contrasted with those of known Galactic SN remnants, 
shown in Fig.~\ref{fig:snr_dist} in the Appendix.

\subsection{Actinides as Secondary Tracers}\label{sec:actinides}

As listed in Table~\ref{tab:nuc}, most of the $\gamma$-ray emitting
nuclei with half-lives in the range
$100$~yr~$\leq t_{1/2}\leq 100$~Myr are actinides with mass number
$226\leq A\leq 250$.  Among these, the most promising candidate for 
an individual remnant search together with $^{126}$Sn (i.e. with a 
comparable half-life) is $^{230}$Th ($t_{1/2}= 7.54(3)\times 10^4$~yr).
$^{230}$Th decays via a long decay chain ending at $^{206}$Pb and
produces a couple of strong $\gamma$-ray lines at sub-MeV energies, e.g.,
351.9~keV (35.6\%) and 609.3~keV (45.5\%).  Despite the somewhat
weaker intensity of these lines, due to its shorter half-life,
$^{230}$Th can generate comparable $\gamma$-ray flux as $^{126}$Sn for
an otherwise similar abundance.

Unfortunately, the production of actinides in NSMs is uncertain
because it requires highly neutron-rich conditions (very low electron
fraction), which may not characterize the bulk of the NSM ejecta.
Nevertheless, one can still estimate the actinide abundance
empirically if one assumes that NSMs must on average produce the
measured solar-$r$ thorium and europium abundances.  In this case, the
number fraction of individual actinide nuclei in the merger ejecta is
roughly given by $Y_{\text{act}}\simeq 3.6\times 10^{-6}$, under the
following assumptions: (i) NSMs occur with a uniform frequency over
the MW's history $\sim 13$~Gyr; (ii) Only actinides between
$226\leq A\leq 250$ produced by NSMs contribute to the solar system
$^{232}$Th abundance, i.e., nuclei heavier than $A=250$ mostly fission
away during and after the $r$-process~\citep{Giuliani:2019oot}.  (iii)
Each NSM produces uniform actinide abundances across the above mass
range.  On the other hand, abundance measurements from metal-poor
stars (those polluted by just a single, or at most a few, $r$-process
events) indicate that the intrinsic Th production yields may vary by a
factor of $\sim 3$ from event to event~\citep{Holmbeck+18,Ji:2018_2}.
Furthermore, theoretical predictions of the actinide abundances based
on nuclear reaction network calculations can vary up to a factor of
$\sim 10$, depending on the assumed nuclear mass models and the
beta-decay half-lives (see e.g.,~\citealt{Holmbeck:2018xet,Eichler:2019rzj}).

Due to these uncertainties, we assess the detection prospects of
actinides in NSM remnants as a function of the actinide abundance.  
We focus on the $^{230}$Th decay line at $609.3$~keV with
$I_g=45.5\%$. Since $^{230}$Th is also the daughter of $^{234}$U
(half-life $t_{1/2}=2.46\times 10^5$~yr), together they produce a
photon number flux:
\begin{align}
\tilde F(d,t)=& \frac{M_{\rm ej} Y_{\rm act} I_g}{4\pi d^2 m_u}
\times [ \frac{1}{\tau_{a}} e^{-t/\tau_{a}} 
 \nonumber \\
& + \frac{1}{\tau_{a}-\tau_{b}} (e^{-t/\tau_{a}}- e^{-t/\tau_{b}})],
\end{align}
where $Y_{\text{act}}$ is the number fraction of both $^{230}$Th and 
$^{234}$U, and $\tau_a$ and $\tau_b$ are the lifetimes of 
$^{230}$Th and $^{234}$U, respectively.

\begin{figure}[htbp!]
  \centering \includegraphics[width=0.9\linewidth]{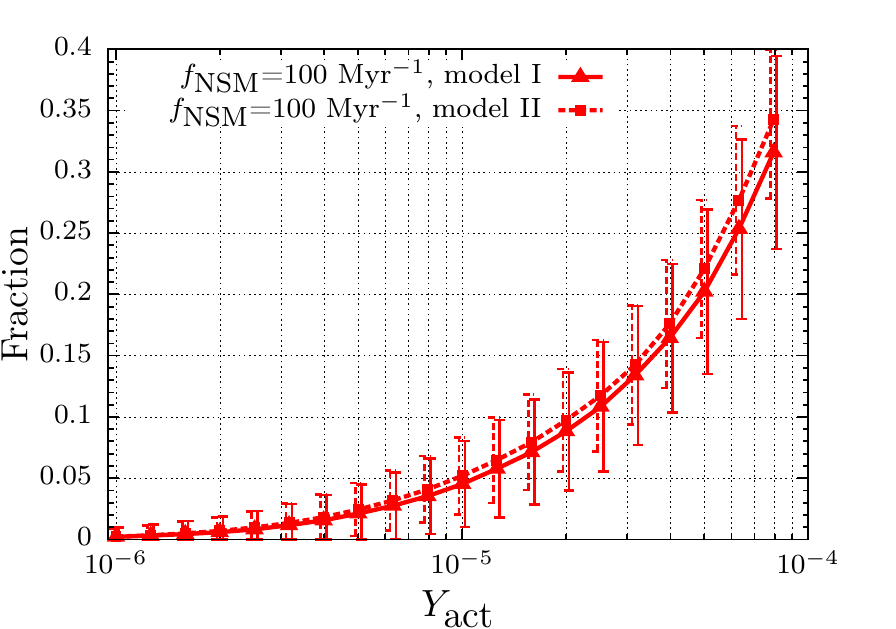}
  \caption{Fraction of NSM remnants that produce $\gamma$-line fluxes
    larger than $F_{\rm lim}=10^{-8}$~$\gamma$~cm$^{-2}$~s$^{-1}$ from
    \emph{both}  the decay of $^{126}$Sn and $^{230}$Th as a function
    of the actinide abundance at production, $Y_{\rm act}$, normalized
    to all remnants with $^{126}$Sn fluxes exceeding the same $F_{\rm
      lim}$.  We show results separately for Model I and II, and in
    each case assume a Galactic merger rate $f_{\rm
      NSM}=100$~Myr$^{-1}$.   
\label{fig:probact}}
\end{figure}

Fig.~\ref{fig:probact} shows the fraction of NSM remnants, considered 
among the population with $^{126}$Sn $\gamma$-ray fluxes [$F(d,t)$] above 
the assumed detection limit $F_{\rm lim}=10^{-8}$~$\gamma$~cm$^{-2}$~s$^{-1}$,
which also produce $^{230}$Th fluxes [$\tilde F(d,t)$] above the same 
value $F_{\rm lim}$, as a function of $Y_{\rm act}$.  
We show separately the results for Model I and II, both assuming 
$f_{\rm NSM}=100$~Myr$^{-1}$.  The fraction satisfying both 
$F > F_{\rm lim}$ and $\tilde F > F_{\rm lim}$ grows roughly 
linearly with $Y_{\rm act}$ within this abundance range.  
Therefore, by measuring this fraction (e.g.~the number of NSM remnants 
which are detectable just by their $^{126}$Sn lines to those also 
detectable through their $^{230}$Th lines), one could in principle 
determine the average actinide production yield, $Y_{\rm act}$.  
By exploring the distributions of line fluxes (in conjunction with 
independent estimates of the source age), it would also be possible 
to infer the diversity in actinide yields and compare to that inferred 
from the abundances of metal-poor stars.  For instance, given the large 
predicted quantity of low electron fraction tidal-tail ejecta in some NS-BH
mergers compared to that in NS-NS mergers (e.g.~\citealt{Foucart+18}), 
a particularly actinide-rich remnant might implicate a NS-BH merger event.

\section{Diffuse Emission}\label{sec:diffuse}

In addition to the individual $\gamma$-ray sources discussed in the previous 
section, there exists a diffuse $\gamma$-ray line background from 
longer-lived nuclei with lifetimes~\mbox{$\gtrsim 1$}~Myr comparable to or 
exceeding the mixing time of the remnants with the Galaxy ISM 
(bottom panels of Fig.~\ref{fig:distancePDF_ModelI}).
Among the longer-lived nuclei, $^{182}$Hf ($t_{1/2}$=8.9~Myr) is arguably 
the most promising candidate because of its relatively strong line
at $\sim 1.1$~MeV ($I_g=35\%$).  

Taking an abundance of $Y(^{182}$Hf$)=7\times 10^{-6}$ in the NSM ejecta, 
again scaled from the solar $r$-process abundances, we calculate the diffuse 
flux (integrated over all mergers over a 50~Myr time span), as well as 
the flux per solid angle at different sky locations.  
We consider two cases: (1) Model I with $f_{\rm NSM}=100$~Myr$^{-1}$; 
(2) an otherwise similar case but which assumes zero physical offset, 
i.e., that the NSMs directly trace the distribution of Galactic 
stellar mass.\footnote{The latter model may provide a reasonable 
approximation for the spatial distribution of $r$-process production 
in scenarios invoking rare types of SNe, such as the birth of 
rapidly-spinning magnetars \citep{Thompson+04,Winteler+12} or 
collapsars \citep{Siegel+18}, as major $r$-process sources.  
However, in principle the remnants from SNe should more tightly trace 
the locations of current star formation.}  

\begin{figure*}[htbp!]
  \centering 
  \includegraphics[width=0.7\textwidth]{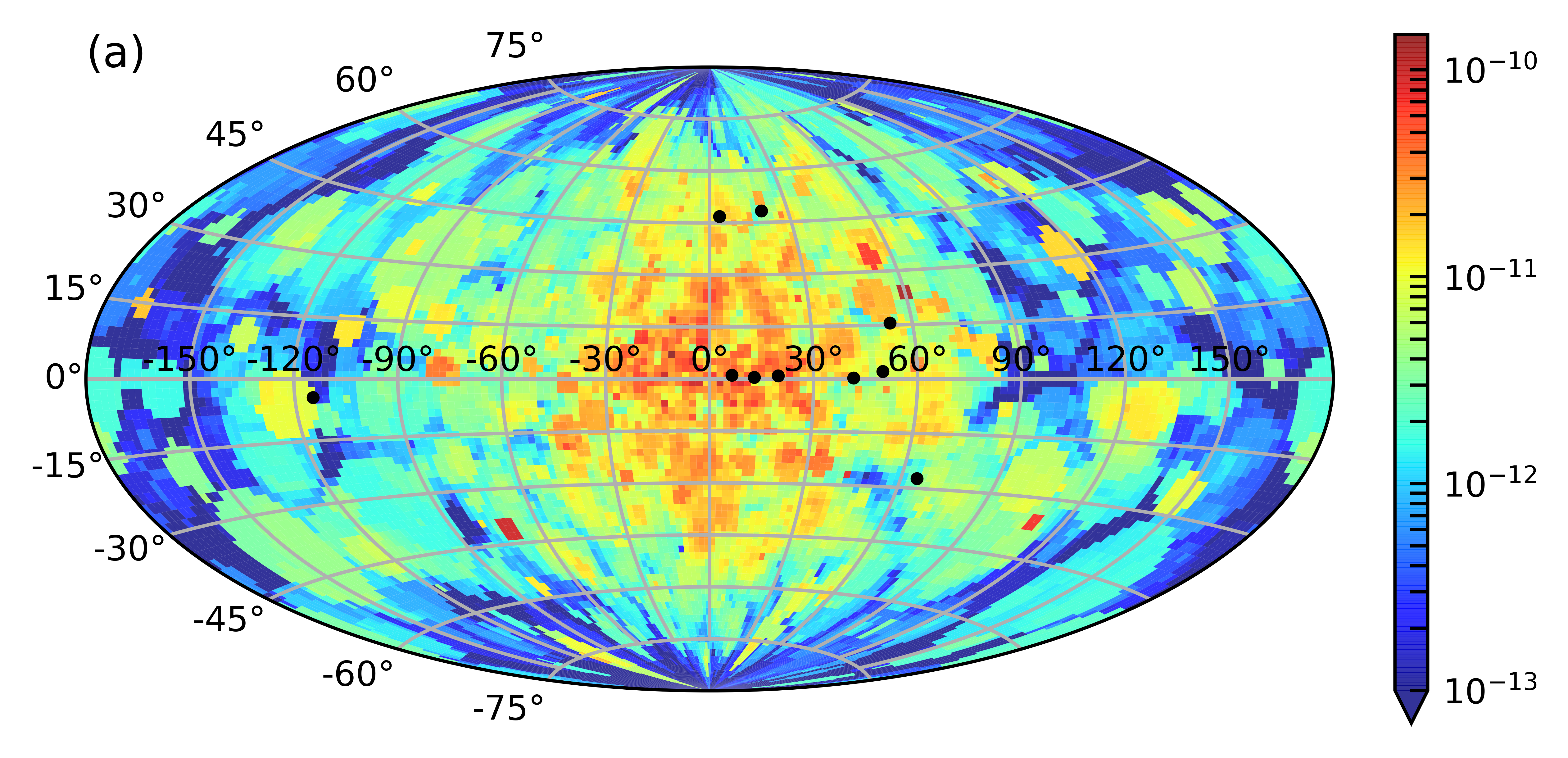}
  \includegraphics[width=0.7\textwidth]{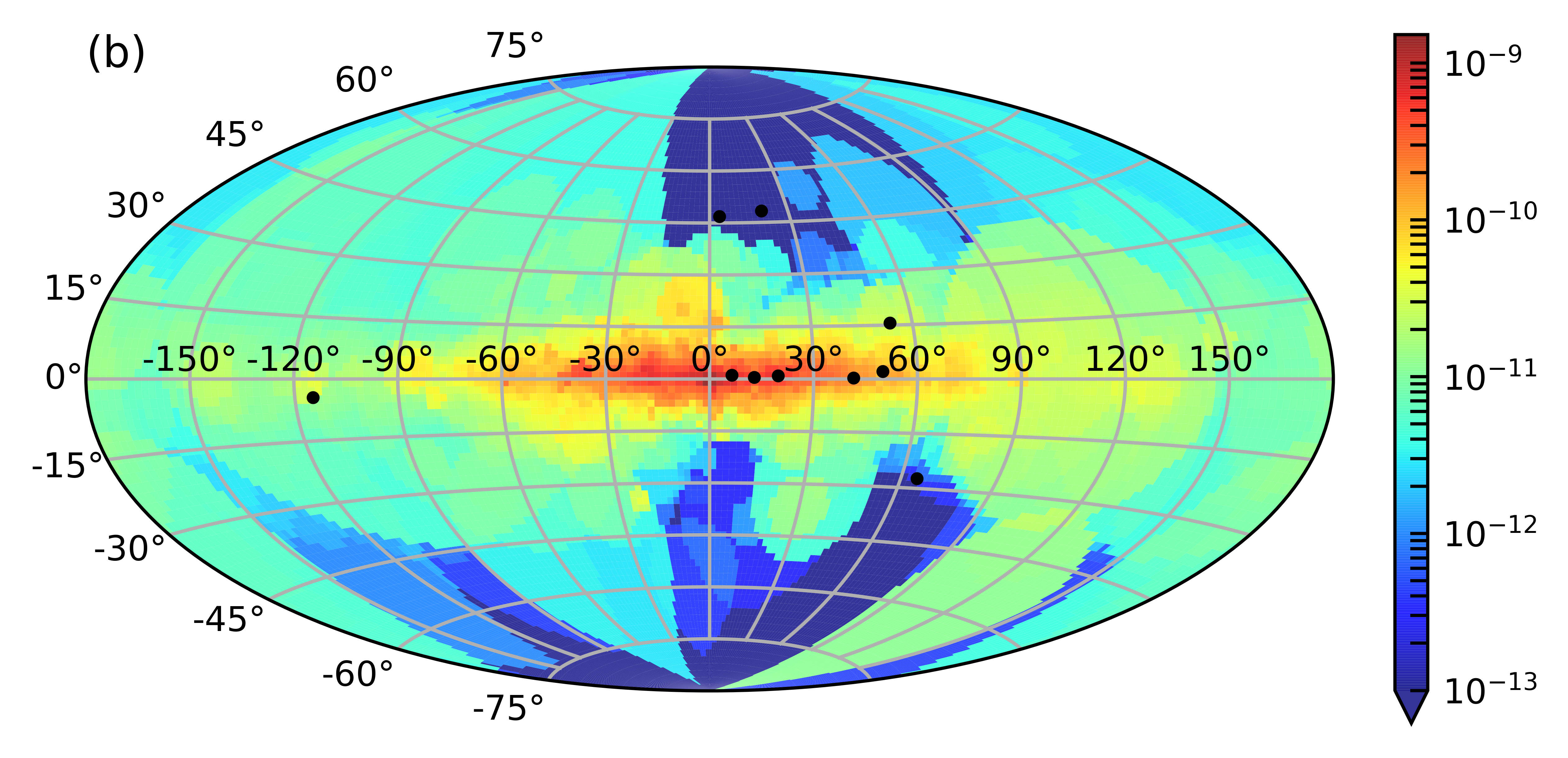}
  \caption{Example realizations of the diffuse flux from the 1.121~MeV 
  decay line of $^{182}$Hf per unit solid angle (in units of 
  $\gamma$~${\rm cm}^{-2}\,{\rm s}^{-1}\, {\rm deg^{-2}}$), shown 
  separately for (a) Model I with frequency of $f_{\rm NSM} = 100$~Myr,
  and (b) with the same frequency but assuming that mergers follow the
  distribution of stars, i.e., no spatial offset due to NS kicks.  
  Black dots show the locations of known double neutron star 
  systems \citep{Lorimer08}, which are primarily concentrated in the 
  Galactic plane.}\label{fig:diffuse}
\end{figure*}

For the first (second) case, we find a total diffuse $^{182}$Hf flux 
integrated over the entire sky of
$\sim 2\times 10^{-8}$($1\times 10^{-7}$)~$\gamma$~cm$^{-2}$~s$^{-1}$.  Fig.~\ref{fig:diffuse} shows, for one realization in each case, the 
flux distribution as a function of solid angle in Galactic coordinates.  
For Model I (top panel), which includes spatial offsets of the NSMs from 
the stars, the diffuse flux extends to large galactic latitudes but 
remains peaked around the GC.  For the zero-offset case (bottom panel), 
the diffuse fluxes instead mostly traces the Galactic plane.  
Unfortunately, in either case the diffuse $r$-process flux is sufficiently 
low $\sim 10^{-10}$~$\gamma$~cm$^{-2}$~s$^{-1}$~deg$^{-2}$ that its 
detection appears beyond the capabilities of even next-generation 
$\gamma$-ray facilities.

Naively, guidance as to the Galactic distribution of $r$-process lines 
could come from observations of $^{26}$Al, which has previously been 
detected through its 1.8 MeV decay line \citep{Pluschke+01,Diehl+06}.  
High energy resolution $\gamma$-ray spectroscopy reveals that $^{26}$Al 
rotates in the same general sense as the Galaxy but has large velocities 
in comparison to other components of the ISM.  \citet{Krause:2015bta} 
explore the hypothesis that this behavior can be explained from the 
kinematics of ``superbubbles" created by the correlated SN explosions 
mixing the ejecta into the hot phase of the ISM.  By contrast, the delay 
time of NSM should result in uncorrelated activity and therefore the 
kinematics could be substantially different.  The eventual measurement 
of such spatial and kinematic properties could therefore help distinguish 
NSMs from SN sources of the $r$-process.  

\section{Search Strategies}
\label{sec:search}
This section addresses general strategies to discover or confirm Galactic
$r$-process $\gamma$-ray line sources.

\subsection{Individual Known Remnants}  
One strategy to discover $r$-process line emission is to search individual 
known core-collapse SN remnants.  
Although ordinary SNe are no longer considered promising sources for the 
heaviest $r$-process elements, they might in some cases produce lighter
$r$-process nuclei extending up to the second peak.  
Furthermore, certain rare classes of SNe (e.g. those which produce
strongly-magnetized neutron stars;
e.g.~\citealt{Thompson+04,Metzger+07,Winteler+12,Mosta+18}) might eject
sufficiently neutron-rich material to create even the heavier $r$-process 
nuclei, similar to a NSM.  Finally, one cannot exclude the possibility 
that a small number of suspected ``SN remnants" are in fact NSM remnants, 
as the two could appear similar at late times in the remnant evolution.

We compute the expected $^{126}$Sn decay line fluxes from all known 
SN remnants in the catalog 
of \citet{Ferrand&SafiHarb12}\footnote{Updated data in \url{http://www.physics.umanitoba.ca/snr/SNRcat/}}, 
normalized to a scenario in which each contains an $r$-process mass 
of $M_{\rm r}\sim 0.01$~$M_\odot$ (i.e.~a $^{126}$Sn mass of 
$\sim 2.14\times 10^{-4}$~$M_\odot$ assuming solar $r$ abundances).  
As compiled in Table~\ref{tab:SNR_flux}, we find 18 remnants that, 
depending on their uncertain distances and ages, could produce 
666.3~keV line fluxes larger than 
$10^{-6}(M_{\rm r}/10^{-2}M_{\odot})$~$\gamma$~cm$^{-2}$~s$^{-1}$.  
Particularly interesting are those cases located outside the Galactic 
plane and without clear evidence for a NS compact object, and which 
therefore could in principle be NSMs masquerading as SN remnants.  
In addition, remnants without clear evidence of X-ray lines associated with 
$\alpha$-elements (O, Mg, Si, S) or Fe (e.g.~as formed in SNe but not NSMs) may be worth targeting.

Also of potential interest are remnants containing magnetar compact 
objects, as these highly-magnetized neutron stars could be formed in 
MHD supernovae (e.g.~\citealt{Winteler+12,Mosta+18}) with an appreciably 
different proto-neutron star wind (e.g.~\citealt{Thompson+04,Metzger+07,Vlasov+17}), which could make 
them more promising $r$-process sources than SNe producing 
less-magnetized NSs.  
Moreover, some magnetars may in fact be born from the NSM~\citep{Metzger:2007cd,Xue:2019nlf},
although the fraction is likely to be only
a few percent \citep[e.g.,][]{Margalit:2019dpi}.
We calculated the $^{126}$Sn decay line fluxes 
from known magnetars listed 
in~\citet{Olausen_Kapsi_2014}\footnote{Data taken from \url{http://www.physics.mcgill.ca/~pulsar/magnetar/main.html}} 
using the estimated distances and their characteristic ages.  
We find that in all cases the line fluxes of magnetar remnants are smaller
than $10^{-6}$~$\gamma$~cm$^{-2}$~s$^{-1}$.  
Among these, SGR~0501+4516 has a possible association
with the SN remnant HB9.  However, its predicted flux is only
$\sim 4\times 10^{-7}$~$\gamma$~cm$^{-2}$~s$^{-1}$, 
which is smaller than that from the HB9, due to the estimated 
distance of $\sim 2$~kpc for the magnetar SGR~0501+4516, versus the 
smaller lower bound on the distance range 0.4--1.2~kpc for the HB9 
remnant (Table~\ref{tab:SNR_flux}).

A complete list of the SN remnants/magnetars and their predicted fluxes 
can be downloaded from~\protect\citet{flux_data}.

\subsection{Survey of Plane and Bulge}
\label{sec:survey}

The most recent Galactic NSMs are likely not associated with known SN
remnants.  While many planned future $\gamma$-ray satellites have all-sky 
monitors that will image the entire sky, for pointed instruments the 
discovery of NSM remnants may therefore require a systematic search of 
the Galactic plane or bulge for the $r$-process $\gamma$-ray line emission. 
To have a high probability of detecting at least a single remnant, 
the search should reach a sensitivity of 
$\lesssim 3\times 10^{-8}$~$\gamma$~cm$^{-2}$~s$^{-1}$ (Fig.~\ref{fig:prob}).  
The predicted angular sizes of the remnants are several degrees
(Fig.~\ref{fig:dist_detrem}), comparable to the angular resolution of
Compton $\gamma$-ray telescopes.  The search should ideally extend to
Galactic latitudes $\pm 20^{\circ}$ (Fig.~\ref{fig:dist_detrem}).  
A sizable fraction ($\sim30\,\%$) of the detectable merger remnants come 
from the direction of the Galactic bulge.  
The astrophysical background in this region is greater, but the ability to
integrate longer on this single region might overcome some of this deficiency.  

Although the most recent Galactic NSM remnant is likely of age $\gtrsim 10^{4}-10^{5}$ yr, it could be substantially younger.  The probability of a NSM of age 
$\lesssim 1$~kyr is $\lesssim 10\%$, but such a remnant (or an equally young $r$-process enriched SN remnant) would have substantially higher $\gamma$-ray luminosity than predicted for $^{126}$Sn and $^{230}$Th, due to enhanced contributions from
shorter-lived actinide nuclei (e.g.~those given in the first few rows in Table \ref{tab:nuc}).

The old NSM remnants of interest will be expanding into 
the ISM at velocities of $\lesssim $ 3000~km s$^{-1}$
(Fig.~\ref{fig:dist_detrem}).  This implies that the Doppler broadening 
of offset of the $\gamma$-ray lines will be relatively modest, at the 
level of $\Delta E/E \lesssim v_{\rm exp}/c \lesssim 1\%$ or less 
(if the $r$-process ejecta is concentrated in the low velocity center 
of the remnant).  Similar or smaller offsets in the energy of the line 
center are expected from the center-of-mass motion of the remnants 
due to the rotational velocity of the MW or their motion relative to the 
disk due to natal supernova kicks.  A $\gamma$-ray telescope with percent- 
or sub-percent level energy resolution would therefore be needed to resolve 
these line features.

\subsection{X-ray Confirmation of Candidates}

In addition to $\gamma$-ray lines from distinct isotopes (e.g.~$^{230}$Th), 
another way to confirm $r$-process remnants is with follow-up observations 
by X-ray satellites \citep{Ripley+14}. 
Our most promising isotope, $^{126}$Sn, contains X-ray L and K lines
centered around 4~keV and 28~keV with summed intensities of $\sim 10$\% and 
30\%, respectively. For sources with 666.3~keV $\gamma$-ray line flux 
exceeding $\sim 2\times 10^{-7}$~$\gamma$~cm$^{-2}$~s$^{-1}$, the 
resulting predicted X-ray line strength at $\sim 4$~keV and 28~keV would 
be $\gtrsim 10^{-16}$~erg~cm$^{-2}$~s$^{-1}$ 
and $\gtrsim 2\times 10^{-15}$~erg~cm$^{-2}$ s$^{-1}$.
Although unlikely to be detectable with current X-ray telescopes 
(e.g.~XMM, NuSTAR, or NICER), our preliminary estimation indicates that they 
may possibly be within the reach of future high-sensitivity missions such 
as eXTP~\citep{Zhang:2016ach}, STROBE-X~\citep{Ray:2018dlb}, and Athena~\citep{Athena2013} for the $\sim 4$~keV line
and HEX-P~\citep{HEXP_Madsen18} for the $\sim 28$~keV line.

\begin{table*}[htb]
  \caption{Ages, distances, and predicted $^{126}$Sn $\gamma$-ray line 
  fluxes at 666.3~keV for nearby SN remnants for which the latter range 
  exceeds $10^{-6}$~$\gamma$~cm$^{-2}$~s$^{-1}$.  We assume an ejecta mass 
  $M_{\rm ej}=0.01$~$M_\odot$ and $Y(^{126}$Sn) $ = 1.7\times 10^{-4}$.  
  The final column indicates the possible association of a compact object 
  (P, M, CCO, PWN denote ``pulsar'', ``magnetar'', ``central compact object'', 
  and ``pulsar wind nebula'', respectively).
  }\label{tab:SNR_flux}
    \begin{ruledtabular}
    \begin{tabular}{ccccc}
        Source & Age ($10^{3}$ yr) & Distance (kpc) & Line Flux
        ($10^{-6}~\gamma~{\rm cm}^{-2}{\rm s}^{-1}$)  & Compact Object or PWN? \\
        \hline
Lupus Loop     & 15--31 & 0.15--0.5 & 5.80--67.60  & P? \\
Vela           & 9--27  & 0.25--0.3 & 16.30--24.78 & P \\
Antlia         & (1--6)$\times 10^3$ & 0.06--0.34 & 0-21.75 & P? CCO? \\
HB9            & 4--7  & 0.4--1.2 & 1.08--9.83 & M? \\
Vela Jr        & 2.4--5.1  & 0.5--1.0 & 1.57--6.32 & CCO? P?\\
3FGL J2014.4+3606 & 11--12 & 0.5--4 & 0.10--6.16 & -- \\
Cygnus Loop    & 10--20 & 0.576--1 & 1.50--4.65 & PWN? \\
Monoceros Loop & 30--150 & 0.6--1.98 & 0.26--4.04 &  ? \\
IC443          & 3--30 & 0.7--2 & 0.36-- 3.22 &  ? \\
2FGL J2333.3+6237 & 7.7     & 0.7    &  3.17 & P? \\
HB21           & 4.8--15 & 0.8--2.1 &  0.34--2.45 & -- \\
G65.3+5.7      & 20 & 0.8 &  2.34 & P? \\
RX J1713.7-3946 & 1-2.1 & 1 &  1.58--1.59 & CCO? \\
DA 495   & 7-155 & 1--3.6 & 0.08--1.56 & PWN? \\
G107.5-01.5 & 3-6  & 1.1 & 1.29--1.30 & -- \\
CTA 1  & 13 & 1.1--1.7 & 0.53--1.26 & P \\
S147 Sh2-240   & 26--34 & 1.1--1.5 & 0.64--1.22 & P \\
R5     & 20--30 & 1.15  & 1.10--1.13 & --
    \end{tabular}
    \end{ruledtabular}
\end{table*}

\subsection{Merger remnants or SN remnants?}
The detection of the decay $\gamma$-ray or X-ray lines 
from the $r$-process nucleus $^{126}$Sn or actinides
does not directly imply that the emitting source is
a NSM remnant. As discussed before, it may in fact 
belong to a rare type of SNe. 
Below we propose several ways to differentiate 
the two scenarios.

First, although Fig.~\ref{fig:dist_detrem} shows
that the NSM remnants preferentially sit toward the direction
of the GC, their distributions in Galactic latitude and vertical
height show striking differences when compared with those from known
SN remnants (see Fig.~\ref{fig:snr_dist}).
Due to the offsets from their birth sites, most of the merger remnants
($\sim 80\%$) have vertical heights $|z|\gtrsim 1$~kpc, while nearly
all SN remnants are located within the MW disk with $|z|\lesssim 0.2$~kpc. 
Thus, for merger remnants, only $\sim 15\%$ are expected to be
within a latitude of $\pm 5^\circ$, while one expects $\gtrsim 90\%$
of SN remnants to be within this latitude (see also
Fig.~\ref{fig:diffuse} which illustrates such difference in terms
of diffuse sources).
Consequently, if a detected remnant has large latitude or vertical distance from the disk, it can be identified unambiguously as
a remnant associated with a NSM.
This fact also highlights again that if future missions can detect
$^{126}$Sn decay $\gamma$-rays lines from $\gtrsim 10$~remnants
(see Fig.~\ref{fig:prob}), it will likely be able to tell us whether
NSMs or rare SNe are the dominant sites of $r$-process nucleosynthesis in the recent history of the MW.

For remnants found to be within the MW disk, 
if they happen to be nearby, $\lesssim 2$~kpc, the
1.8~MeV $\gamma$-ray line flux from the decay of $^{26}$Al can be
larger than $\sim 10^{-7}$~$\gamma$~cm$^{-2}$~s$^{-1}$,
for an assumed typical yield of $^{26}$Al mass of 
$\sim 5\times 10^{-5}$~$M_\odot$ from a SN
(e.g., \citet{Limongi:2006xx,Tur:2010xx,Sieverding:2018rdt}).
As the production of $^{26}$Al in NSM should be negligible,
a co-detection of $^{26}$Al decay line, together with 
the lines from $^{126}$Sn can, therefore, be used to indicate
whether a close-by remnant is resulting from NSM or SN. 
Likewise, potential 1.17 \& 1.33~MeV $\gamma$ lines from 
$^{60}$Fe$\rightarrow ^{60}$Co$\rightarrow ^{60}$Ni,
or the $\sim 3.3$~keV (5.4~keV) X-ray lines from the decay
of $^{41}$Ca ($^{53}$Mn), if identified, will
similarly indicate that the emitting source is not a NSM remnant.
Moreover, X-ray signature at 0.5--10~keV due to the presence of
$\alpha$-elements and/or the iron group in a young remnant~\citep{vink:2012xx}
before entering the Sedov-Taylor phase will certainly rule
out the remnant being associated with a NSM.

\section{Discussions and conclusions}\label{sec:disconcl}

Detecting $r$-process $\gamma$-ray line emission from extremely young
extragalactic NSMs (e.g. those discovered by LIGO) is likely to be 
challenging for the foreseeable future given their large distances
(e.g.~\citealt{Hotokezaka+16,Li:2019xx,Korobkin:2019uxw}).
Motivated by earlier work \citep{Qian+98,Qian+99,Ripley+14} on $\gamma$-ray and X-ray line 
emission from the MW $r$-process sources, as well as
the measured rate of NSMs by LIGO/Virgo Collaboration and the $r$-process 
yield inferred from GW170817,
we have shown that the detection of $^{126}$Sn decay is possible from the most recent 
NSM remnants in our Galaxy for future MeV $\gamma$-ray telescopes 
which reach photon line flux sensitivities of 
$\lesssim 10^{-7}-10^{-6}$~$\gamma$~cm$^{-2}$~s$^{-1}$.  
Furthermore, the co-detection of decay lines from $^{230}$Th 
in such remnants would constrain the uncertain yield of actinides, 
a question which remains open in spite of the kilonova detection 
from GW170817 \citep{Wanajo18,Zhu+18,Wu+19}.  
Detection of the diffuse line background from $^{182}$Hf could in 
principle map out the spatial distribution of $r$-process sources 
over the past $\lesssim 10$~Myr, allowing for another test of NSM 
versus SN origin based on the vertical extent of this emission above 
the Galactic plane.  However, the low level of the diffuse flux 
$\lesssim 10^{-10}$~$\gamma$~cm$^{-2}$~s$^{-1}$ deg$^{-2}$ will be 
a challenging target even for next generation $\gamma$-ray telescopes.

Arguably the biggest challenge in detecting $\gamma$-ray emission from 
NSM remnants is finding their locations in the first place.  
A wide-field search would naturally be conducted for future planned 
all-sky monitors, such as AMEGO, COSI, e-ASTROGAM, GRAMS, and LOX.  
However, for pointed telescopes with narrower fields of view, the 
remnant population would most easily be discovered via a systematic 
search of the Galactic plane prioritized toward the GC/bulge, as we 
predict most detectable remnants to reside within $\pm 20^\circ$ of the GC.
Moreover, since a very small fraction of nearby ``SN remnants" could in 
fact be NSM remnants, or arising from rare classes of explosions that 
synthesize heavy $r$-process elements (particularly those supernovae 
creating magnetars), pointed searches for $^{126}$Sn $\gamma$-lines 
from these remnants are promising initial test targets for next-generation
$\gamma$-ray detectors.
We also note that although the line flux integrated over a large 
angular area can be much higher than the fluxes from individual sources, 
(e.g., $\sim 2$--$4\times 10^{-6}$~$\gamma$~cm$^{-2}$~s$^{-1}$ for the
$^{126}$Sn 666.3~keV line in Model I with $f_{\rm{NSM}}=100$~Myr$^{-1}$ when
integrated over the entire sky),
given the larger background, the detection of individual sources may in fact 
be more favorable.

Several uncertainties affect our predictions.  First, we have assumed
that each merger produces an $r$-process abundance distribution
following the Solar $r$ abundances for nuclei above mass number
$A=90$.  However, detailed analysis of the time evolution of the
kilonova spectrum from GW170817 indicated that it contains a
lanthanide mass fraction of only $X_{\rm lan}\sim 10^{-2}$~(e.g.,~\citealt{Kasen+17,Kawaguchi+18}), roughly an order of magnitude smaller than that of the solar $r$ abundance pattern above $A=90$.  Future kilonova
observations, together with the improved theoretical modeling, as well
as the potential late-time detection, may be able to tell whether most
of the NSMs produce primarily elements above $A\geq 90$ with a similar
amount above the second peak as of the Solar $r$ abundance
distribution.
If the $^{126}$Sn yield in NSM is smaller than what we have
assumed, then the prospects of detecting their $\gamma$-rays would
obviously be correspondingly weaker.  On the other hand, one may
utilize the fact that $^{126}$Sn has three strong lines with clearly
predicted energies and branching ratios, to increase the detection
confidence in marginal cases.

Our model for the spatial offset distribution of NSM remnants is 
motivated empirically by observations of short gamma-ray bursts,
which are compatible with average center-of-mass velocities of the binary systems of $\sim 20$--140~km~s$^{-1}$.
Alternatively, the NSM offsets from their birth site can be modeled on a more physical basis by
folding a realistic distribution of SN kick, mass loss during the second SN, stellar motions in different 
part of MW, the motion of the NS binary under the influence of the MW potential, in addition to the evolution of massive stars in binaries
(e.g., \citealt{Bloom:1998zi,Belczynski:2006br,Chruslinska:2017odi, Andrews:2019ome}). This is beyond the scope of this paper. We note however, that such an approach would necessarily involve large theoretical
uncertainties at present but may be improved in the future.
For roughly 2/3 of the currently known sample of binary neutron star systems in the MW, the second neutron star appears to have received low kick and low mass ejection, such that they would have negligible change in their center-of-mass velocity.  These systems would thus be confined close to the disk and will merge near the Galactic plane (e.g.~\citealt{Beniamini&Piran16,Tauris+17}). Nevertheless, the remaining $\sim$ one-third of binary neutron star systems appear to have received relatively high kick velocities (and consequently a large change in center-of-mass velocity).  Note that currently it is not possible to ascertain whether the small observed sample is representative of the total population of Galactic neutron star binaries due to the limited sample size and unknown selection bias.  The detection of $\gtrsim 10$ NSM remnants with future $\gamma$-ray mission proposed here can provide new insights into these open questions.

We encourage further concept studies of various proposed gamma-ray telescope missions, which utilize
a range of different detector types, in order to assess their abilities to provide sufficient spatial 
and energy resolution, as well as a large effective area to achieve
the required line sensitivities, as needed to detect NSM remnants in the 
Milky Way.  For example, the germanium detector in COSI and  and INTEGRAL-SPI demonstrated an excellent energy resolution.
The double-sided silicon strip detector proposed in AMEGO and 
e-ASTROGAM can provide an excellent spatial resolution as successfully
demonstrated in Fermi-LAT.  A liquid argon time projection chamber (LArTPC) detector proposed in 
GRAMS can be expanded to a larger scale detector considering that 
argon is cost effective and widely used in neutrino and
dark matter search experiments.  A gaseous time projection chamber (TPC) detector in 
the electron tracking Compton camera (ETCC) is particularly optimized to track 
Compton scattered electrons, which can further constrain the direction of
the incoming $\gamma$-ray and reduce coincident background events. The background components on a low-Earth orbit have been well-studied based 
on the previous experiments and modeled in the energy range of 10~keV to 
100~GeV, covering most of the energy range of interest~\citep[see e.g.,][]{Cumani:2019ryv}, but further evaluation will also be needed for each specific mission.

Near the completion of this manuscript, we became aware of independent work on the 
$\gamma$-ray line signals from NSMs, both from extragalactic events and from 
remnants in the Milky Way~\citep{Korobkin:2019uxw}.  While our work 
is more specifically focused on Galactic remnants, our conclusions 
broadly agree with those of these authors.

\acknowledgements
The authors thank Hsiang-Kuang Chang, Roland Diehl, Enrico Ramirez-Ruiz, 
Mohammad Safarzadeh, Thomas Siegert, Anna Watts and an anonymous referee 
for their helpful comments.
MRW acknowledges support from the Ministry of Science and Technology, 
Taiwan under Grant No. 107-2119-M-001-038, and
the Physics Division, National Center of Theoretical Science of Taiwan. 
PB is partly supported by National Natural Science Foundation of 
China (fund \#11533006).
BDM acknowledges support from NASA through the Astrophysics Theory 
Program (grant \#NNX16AB30G).
GMP is partly supported by the Deutsche Forschungsgemeinschaft 
(DFG, German Research Foundation) - Projektnummer 279384907 - SFB~1245.
TA is supported and funded by Department of Energy (DE-AC02-76SF00515).
EB is supported by an appointment to the NASA Postdoctoral Program at 
the Goddard Space Flight Center, administered by Universities Space
Research Association under contract with NASA.
CJH acknowledges support from NASA under Contract No.~NNG08FD60C.
JB is supported by the National Aeronautics and Space
Administration (NASA) through the Einstein Fellowship
Program, grant number PF7-180162.
GK acknowledges support from the National Science Foundation (grant number PHY-1404209). 
MRW and PB thank the Yukawa Institute for Theoretical Physics in
Kyoto for support in the framework of the YITP-T-18-06 workshop,
during which several aspects of this work have been discussed. 

\appendix
\section{Long-lived nuclei and their decay $X$- and $\gamma$-ray lines}\label{sec:nucline}

Table~\ref{tab:nuc} lists the decay channel, half-life $t_{1/2}$, 
major lines with intensity larger than $30\%$ for all 22 $r$-process 
nuclei that produce decay $\gamma$-lines with half-lives between 
$100$~yr$\leq t_{1/2}\leq $ 100~Myr. Except for $^{126}$Sn, $^{129}$I,
and $^{182}$Hf, all other 19 isotopes  are actinides.

\begin{figure}[htbp!]
  \centering \includegraphics[width=\linewidth]{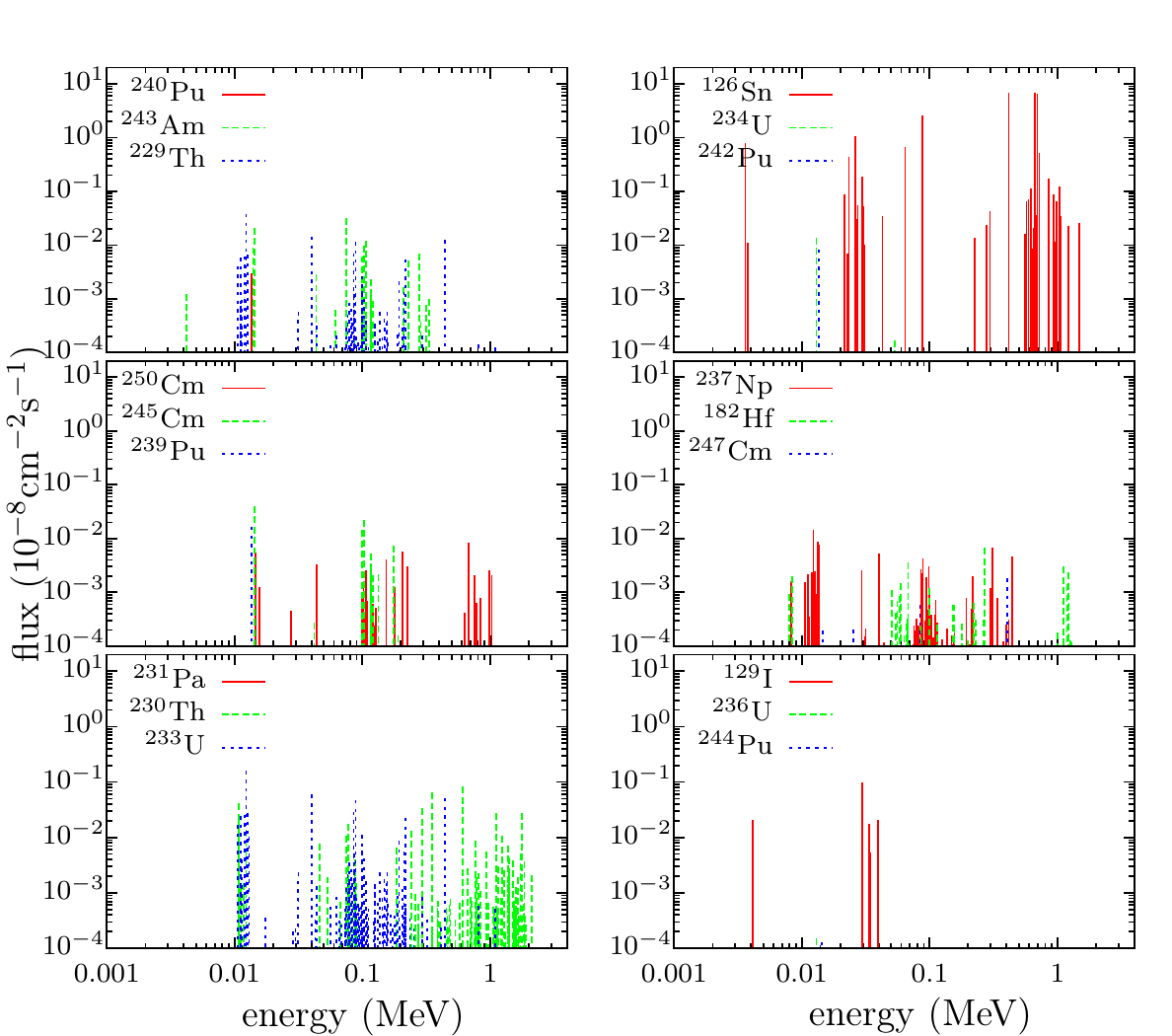}
  \caption{Spectrum of $X$- and $\gamma$-ray line fluxes from a NSM 
  remnant of distance $9$~kpc and age $5\times 10^{4}$ yr (characteristic 
  of the youngest Galactic remnants), including most nuclei from
  Table~\ref{tab:nuc}.   
  The abundance of each isotope is determined as described in the text.  
  Note that the first four nuclei in Table~\ref{tab:nuc} with short 
  half-lives $t_{1/2}<0.002$~Myr do not contribute fluxes above 
  $10^{-12}$~$\gamma$~cm$^{-2}$~s$^{-1}$ and therefore are not included.
  \label{fig:lines}}
\end{figure}

In Fig.~\ref{fig:lines}, we show the line fluxes produced from the decay 
of those nuclei listed in Table~\ref{tab:nuc} at times much shorter 
than $100$~yr. We assume that a NSM at $9$~kpc produces an amount
of ejecta $M_{\rm ej}=0.04$~$M_\odot$. The abundance of $^{126}$Sn, 
$^{129}$I, and $^{182}$Hf inside the ejecta are assumed to follow the 
solar $r$ pattern in the same way described in the main text.
For the actinides, we assume again, a number fraction of 
$Y_{\rm act} = 3.6\times 10^{-6}$.

\section{Galactic SN Remnants}
\begin{figure}[htbp!]
  \centering 
  \includegraphics[width=0.8\textwidth]{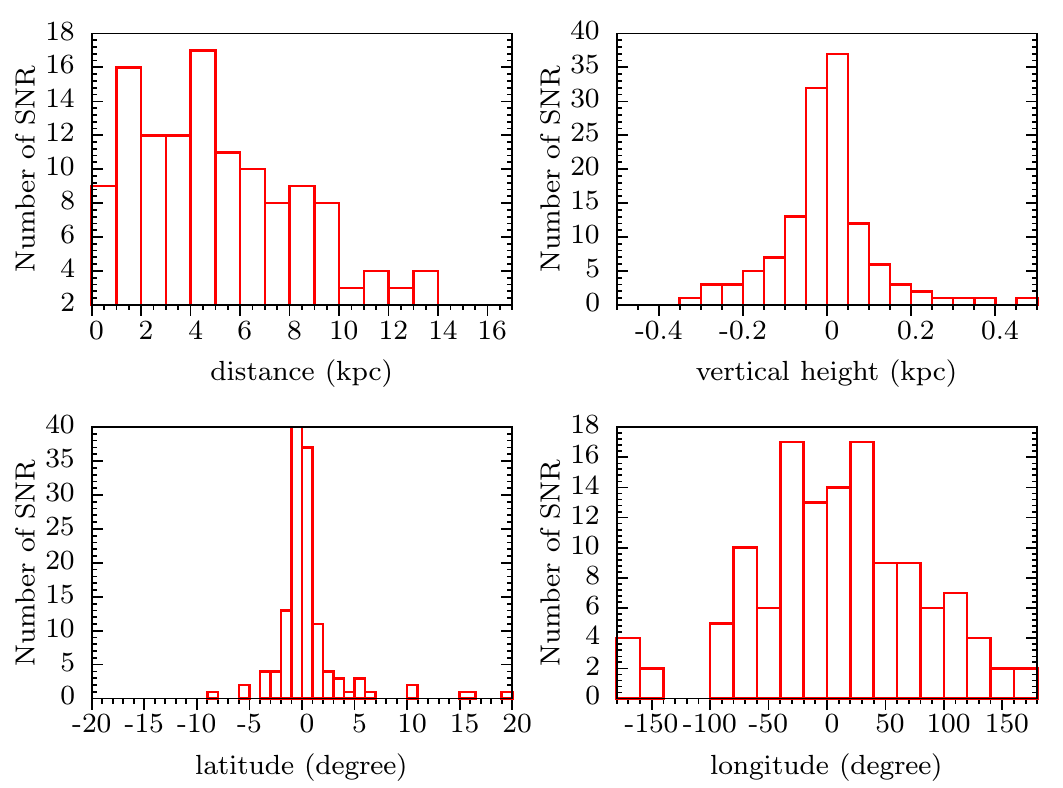}
  \caption{Distribution of the distances, vertical height,  
  latitude, and longitude (in the Galactic coordinate), for known SN remnants. 
  Data taken from~\citet{Ferrand&SafiHarb12}.
  }\label{fig:snr_dist}
\end{figure}

Fig.~\ref{fig:snr_dist} shows the distribution of distances, vertical 
height above the mid-plane of the MW-disk, the Galactic latitude, and 
longitude for all known SN remnants (excluding Type Ia SNe).  
This clearly shows that all the SN remnants are located within 
$\sim \pm 0.5$~kpc from the mid-plane of the Galactic disk and small 
latitudes within $\sim\pm 5^\circ$, in contrast to the predicted NSM 
remnant distribution (see Fig.~\ref{fig:dist_detrem}).  
Nevertheless, as mentioned in the main text, it remains possible that 
some a few SN remnants are in fact from NSMs.

\begin{table*}[ht]
  \caption{List of 22 nuclei that produce decay $X$- and
    $\gamma$-lines with half-lives $100$~yr~$\leq t_{1/2}\leq $
    100~Myr.  For lines produced via shorter-lived nuclei inside the
    decay sequences from a parent isotope, the name of those nuclei
    are given inside the parenthesis next to the line energy.
    Note that for each nucleus, only the major lines with line
    intensities larger than $30\%$, or the strongest lines (if there is no line with intensity larger than 30\%), are listed.
    The table is compiled from data listed in the NuDat 2
    database~\protect\citep{Nudat2}.}\label{tab:nuc}
  \begin{ruledtabular}
    \begin{tabular}{ccccc}
Isotope & Decay channel & $t_{1/2}$ & major lines\tablenotemark{a} & intensity\\ 
        &               & ($10^5$~yr)      & (keV) & $\geq 30\%$ \\
  \hline
 $^{249}$Cf & $\alpha$ to $^{245}$Cm        & 0.0035 & 388 & 66.0 \\ \hline
 $^{241}$Am & $\alpha$ to $^{237}$Np        & 0.0043 & 13.9 & 37.0 \\ 
 & & & 59.5 & 35.9 \\ \hline
      $^{251}$Cf & $\alpha$ to $^{247}$Cm        & 0.0090 & 15 & 53.0 \\ \hline
 $^{226}$Ra & $\alpha\beta$ to $^{206}$Pb  & 0.016 & 351.9 ($^{214}$Pb) & 35.6 \\ 
    & & & 609.3 ($^{214}$Bi)& 45.5 \\ \hline
 $^{240}$Pu & $\alpha$ to $^{236}$U        & 0.066 & 13.6 & 9.6 \\ \hline
 $^{243}$Am & $\alpha\beta$ to $^{239}$Pu        & 0.074 & 14.3 ($^{239}$Np)& 43.3 \\ 
 & & & 74.66 & 67.2 \\ \hline
 $^{229}$Th & $\alpha\beta$ to $^{209}$Bi  & 0.079 & 12.3 &  80.0\\
 & & & 40.0 ($^{225}$Ra)& 30.0 \\ \hline
 $^{250}$Cm & $\alpha\beta$ to $^{246}$Cm        & 0.083 & 679.2 ($^{246}$Am) & 11.5 \\ \hline
 $^{245}$Cm & $\alpha\beta$ to $^{237}$Np        & 0.084 & 14.3 & 53.0 \\ \hline
 $^{239}$Pu & $\alpha$ to $^{235}$U        & 0.24 & 13.6 & 4.3 \\ \hline
 $^{231}$Pa & $\alpha\beta$ to $^{207}$Pb & 0.33 & 12.7 & 45.0 \\ \hline
 $^{230}$Th & $\alpha\beta$ to $^{208}$Pb   & 0.75 & 351.9 ($^{214}$Pb) & 35.6 \\
    & & & 609.3 ($^{214}$Bi)& 45.5 \\ \hline
 $^{233}$U & $\alpha\beta$ to $^{209}$Bi  & 1.59 & 12.3 ($^{229}$Th) & 80.0 \\
 & & & 40.0 ($^{225}$Ra) & 30.0 \\
 \hline
 $^{126}$Sn & $\beta$ to $^{126}$Te & 2.3 & 87.6 & 37.0 \\
 & & & 414.7 ($^{126}$Sb) & 98 \\ 
 & & & 666.3 ($^{126}$Sb) & 100\\ 
 & & & 695.0 ($^{126}$Sb) & 97\\ \hline 
 $^{234}$U & $\alpha$ to $^{230}$Th & 2.46 & 13.0 & 10.0 \\ \hline
 $^{242}$Pu & $\alpha$ to $^{238}$U & 3.73 & 13.6 & 8.6 \\ \hline
 $^{237}$Np & $\alpha\beta$ to $^{209}$Bi  & 21.4 & 12.3 ($^{229}$Th)& 80.0 \\
 & & & 13.3 & 49.3 \\
 & & & 40.0 ($^{225}$Ra)& 30.0 \\
 & & & 311.9 ($^{233}$Pa)& 38.5\\ \hline
 $^{182}$Hf & $\beta$ to $^{182}$W & 89 & 67.7 ($^{182}$Ta)& 42.6 \\ 
 & & & 270.4 & 79.0 \\ 
 & & & 1121.3 ($^{182}$Ta)& 35.24 \\ \hline
 $^{247}$Cm & $\alpha\beta$ to $^{235}$U  & 156 &  14.3 ($^{239}$Np)& 43.3 \\ 
 & & & 74.66 ($^{243}$Am) & 67.2 \\
 & & & 402.4 & 72.0 \\ \hline
 $^{129}$I & $\beta$ to $^{129}$Xe & 157 & 29.782 & 36  \\ \hline
 $^{236}$U & $\alpha$ to $^{232}$Th & 234 & 13.0 & 9.0 \\ \hline
 $^{244}$Pu & $\alpha\beta$ to $^{236}$U   & 811 & 14.3 ($^{240}$Np) & 27.0  \\
 & & & 554.6 ($^{240}$Np) & 20.9\\
    \end{tabular}
 \tablenotetext{a}{Note that the lines between 10--20~keV are
    labeled as XR~l in the database and are in fact the aggregation
    of all L-shell X-ray lines.}
  \end{ruledtabular}
\end{table*}

\end{document}